\documentclass[twocolumn,preprintnumbers,amsmath,amssymb]{revtex4}

\usepackage{graphicx}
\usepackage{dcolumn}
\usepackage{bm}

\begin{document}

\title{ Surface Partition of  Large Fragments}

\author{K. A. Bugaev$^{1,2}$, L. Phair$^{2}$, J. B. Elliott$^{2}$ and L. G. Moretto$^{2}$}
\affiliation{$^1$Bogolyubov Institute for Theoretical Physics,
Kiev, Ukraine\\
%
%
$^2$Lawrence Berkeley National Laboratory, Berkeley, CA 94720, USA
}

\date{\today}
\begin{abstract}
The surface partition of large fragments is derived analytically 
within a simple statistical model by the Laplace-Fourier 
transformation method. In the limit of  small amplitude  deformations,
 a suggested Hills and Dales Model reproduces 
the leading term of the famous Fisher
result for the surface entropy with an accuracy of a few percent.
The surface partition of finite fragments is discussed as well.
\end{abstract}

\maketitle


\section{Introduction}

During last forty years the Fisher droplet model (FDM) \cite{Fisher:64}, on one hand, was 
extensively used to analyze the condensation  of  gaseous phase (droplets of all sizes)    
into liquid. The gaseous phases are ranging from mixture of  nuclear fragments \cite{Moretto}  
to the various  clusters on  2- and 3-dimensional Ising lattices \cite{Ising:clust}.
On the other hand, the FDM inspired the formulation of a more sophisticated model,
the statistical multifragmentation model (SMM) \cite{Bondorf:95}, which describes not
only the gaseous phase of nuclear fragments, but the liquid phase (nuclear matter) 
as well on the same footing \cite{Bugaev:00,Bugaev:01,Reuter:01}. 

Applying the FDM to the nuclear fragment  of $A$-nucleons, 
one can  cast its  free-energy $F_A$ as follows 
\begin{equation}\label{one}
F_A = - W ~A
+ \sigma (T)~ A^{2/3} + \tau  T\ln A~
\end{equation}
Here $W$ is the bulk binding energy per nucleon,
$\sigma (T)$ is the
temperature dependent surface tension which { about the critical temperature $T_c$}  is parameterized
in the following form:
$
\sigma (T)=\sigma_{\rm o} [ 1~ - ~T/T_c],
$
with $\sigma_{\rm o} \approx 18$~MeV and $T_c=18$~MeV ($\sigma=0$
at $T \ge T_c$). The last contribution in Eq.~(\ref{one}) generates the famous Fisher's term with
dimensionless parameter
$\tau$.
From the study of the combinatorics of lattice gas clusters in two dimensions,
Michael Fisher had postulated Eq. (\ref{one}) and {  this kind of  
temperature dependence} of the surface tension because
the latter  naturally leads to the existence of  critical temperature $T_c$.  
This is, of course, not a unique parametrization of the surface tension.
The SMM, for instance, successfully  employs another one \cite{Bondorf:95}
$
\sigma (T)= \sigma_{\rm o}
[(T_c^2~-~T^2)/(T_c^2~+~T^2)]^{5/4}.
$
Therefore,  it is necessary to study a few simple, but quite  fundamental questions,
``What is the origin of  the Fisher  parametrization for the temperature dependent surface tension in
three dimensions?
Does  any physical motivation favor the Fisher parametrization?'' 
This work is devoted to these questions of fundamental importance.

\section{Hills and Dales Model}

To answer these main questions we will consider the statistical model of surface deformations. 
We will, however, impose a necessary constraint that the deformations should conserve the total volume of 
the fragment of $A$-nucleons. As we will see, the most interesting result corresponds to
the deformations of vanishing amplitude. Therefore, it is clear that the shape  of deformation
cannot be important and we can choose  one which is  regular to  simplify our presentation.
For this reason  we shall consider  cylindrical deformations of positive height $h_k>0$ (hills) 
and negative height $-h_k$ (dales), with  $k$ nucleons at the base. 
For simplicity it is assumed that the top (bottom) of the hill (dale) has the same shape as 
the surface of original  fragment of $A$ nucleons. 
Our main assumptions are as follows:
(i) the statistical weight of deformations $\exp\left( - \sigma_{\rm o} |\Delta S_k|/T /s_1 \right) $ 
is given  by the Boltzmann factor due to the  change of the surface $|\Delta S_k|$ in units of 
the surface per one nucleon $s_1$;
(ii) the hill's heights $h_k \le H_k$ ($H_k$ is the maximal height of the hill with $k$ nucleons at the base)
have the same probability $d h_k/ H_k$ besides the statistical one; 
(iii) the same two assumptions are valid for the dales as well. 
As we will see, 
these assumptions are not too restrictive, but allow us to simplify the analysis.  

Under adopted assumptions it is possible to find the one-particle statistical partition of
the deformation of the $k$ nucleons base as a convolution of two probabilities: 
\begin{equation}\label{two}
z_k^{\pm} \equiv \hspace*{-0.15cm} \int\limits_0^{\pm H_k} \hspace*{-0.15cm} \frac{ d h_k}{ \pm H_k}\,
{\textstyle e^{ - \frac{\sigma_{\rm o} P_k |h_k| }{T s_1} } }
= T s_1 
\frac{\left[1 - {\textstyle e^{ - \frac{\sigma_{\rm o} P_k H_k}{ T s_1} } } \right] }{\sigma_{\rm o} P_k H_k }\,,
\end{equation}
where upper (lower) sign corresponds to hills (dales). Here $P_k$ is the perimeter of the cylinder base.   
Now we have  to find a geometrical partition (degeneracy factor) or the number of ways to place
the center of a  given  deformation on the surface of $A$-nucleon fragment while it is occupied
by the set of $\{n_l^\pm  = 0, 1, 2,...\}$  deformations of the $l$ nucleons base. 
Our next assumption is that the desired geometrical partition
can be given  in the  excluded volume  approximation
\begin{equation}\label{three}
{\cal G} = 
\frac{ S_A - \sum\limits_{k = 1}^{K_{max} } k\, (n_k^+ ~ + ~ n_k^-) \, s_1 }{s_1} \,, 
\end{equation}
where $s_1 k$ is the area  occupied by the deformation of $k$ nucleon base ($k = 1, 2,...$), 
$ S_A$
is the  full surface of the fragment,
and $K_{max} (S_A) $ is the $A$-dependent size of the maximal allowed base on the fragment.   
It is clearly seen now that 
the denominator in the right hand side (r.h.s.) of (\ref{three}) corresponds to the available surface
to place the center of each of $\{n_k^\pm \}$ deformations that exist on the fragment surface.
It is necessary to impose the condition ${\cal G} \ge 0$ which ensures that the deformations
do not overlap on the available surface of the fragment. 
Eq. (\ref{three}) is the Van der Waals excluded volume  
approximation usually used in statistical mechanics at not too high particle densities 
\cite{Bondorf:95,Bugaev:00,Goren:81} and it can be derived for the objects of different sizes in a spirit of
a method  proposed in  \cite{Zeeb:02}.

According to Eq. (\ref{two}) the statistical partition for the hill with a $k$-nucleon base matches
 that of the dale, i.e.   $z^+_k = z^-_k$. 
Therefore,  
the grand canonical surface partition (GCSP)
\begin{equation} \label{four}
Z(S_A)= \hspace*{-0.10cm} \sum\limits_{\{n_k^\pm = 0 \}}^\infty \hspace*{-0.10cm} \left[ \prod_{k=1}^{ K_{max} }
\frac{ \left[ z_k^+ {\cal G} \right]}{n^+_k!}^{n^+_k} \frac{ \left[ z_k^- {\cal G} \right]}{n^-_k!}^{n^-_k}\right]
\Theta(s_1 {\cal G})\, 
\end{equation}
corresponds to the conserved (on average) volume of the fragment because the probabilities 
of hill and dale of the same base are identical. 
The presence of $\Theta(s_1 {\cal G})$-function in (\ref{four}) ensures that only configurations
with positive value of the free surface of fragment are taken into account.  
However, this makes the calculation of the GCSP very difficult. 
Because of the explicit $S_A$ dependence
of the maximal base of  deformations via $K_{max} (S_A)$ the standard trick to deal with
the excluded volume partitions, the usual Laplace transform method 
\cite{Goren:81, Bugaev:00, Bugaev:01} in $S_A$,
cannot overcome this difficulty.
%
%
However, the GCSP (\ref{four}) can be solved with the help of the recently developed
Laplace-Fourier technique \cite{Bugaev:04}. The latter employs the identity 
\begin{equation}\label{five}
G (S_A) =
%
\int\limits_{-\infty}^{+\infty} d \xi~ \int\limits_{-\infty}^{+\infty}
  \frac{d \eta}{\sqrt{2 \pi}} ~
{\textstyle e^{ i \eta (S_A - \xi) } } ~ G(\xi)\,,
\end{equation}
which is based on the Fourier representation of the Dirac $\delta$-function.
The representation (\ref{five}) allows us to decouple the additional  
$S_A$-dependence in $K_{max} (S_A)$  and reduce it to the exponential one,
which can already be integrated by the Laplace transform \cite{Bugaev:04}
\begin{eqnarray} \label{six}
&&\hspace*{-0.5cm}{\cal Z}(\lambda)~\equiv ~\int_0^{\infty}dS_A~{\textstyle e^{-\lambda S_A}}
~{ Z}(S_A) = \nonumber\\
&&\hspace*{-0.5cm}\int_0^{\infty}\hspace*{-0.2cm}dS^{\prime}
\int\limits_{-\infty}^{+\infty} d \xi~ \int\limits_{-\infty}^{+\infty}
\frac{d \eta}{\sqrt{2 \pi}} ~ { \textstyle e^{ i \eta (S^\prime - \xi) - \lambda S^{\prime} } } 
%
\hspace*{-0.1cm}\sum\limits_{\{n_k^\pm = 0 \}}^\infty
\hspace*{-0.1cm} \left[\prod_{k=1}^{K_{max}(\xi)} \right.
\nonumber \\
&&\hspace*{-0.5cm}
%
%
\left.
\frac{ \left[ z_k^+ S^{\prime} {\textstyle e^{ k\,s_1(i \eta -\lambda ) } } \right]}{n^+_k!~ s_1^{n^+_k} }^{n^+_k} 
\frac{ \left[ z_k^- S^{\prime} {\textstyle e^{ k\,s_1(i \eta -\lambda ) } } \right]}{n^-_k!~ s_1^{n^-_k} }^{n^-_k}
\right]
\Theta(S^{\prime}) =
\nonumber \\
&&\hspace*{-0.5cm}\int_0^{\infty}\hspace*{-0.2cm}dS^{\prime}
\int\limits_{-\infty}^{+\infty} d \xi~ \int\limits_{-\infty}^{+\infty}
 \frac{d \eta}{\sqrt{2 \pi}} ~ { \textstyle e^{ i \eta (S^\prime - \xi) - \lambda S^{\prime}
+ S^\prime {\cal F}(\xi, \lambda - i \eta) } }\,.
%
%
%
\end{eqnarray}
After changing the integration variable 
$S_A \rightarrow S^{\prime} = S_A - \sum\limits_{k = 1}^{K_{max}(\xi) } k\, (n_k^+ ~ + ~ n_k^-) \, s_1 $,
the constraint of $\Theta$-function has disappeared.
Then all $n_k$ were summed independently leading to the exponential function.
Now the integration over $S^{\prime}$ in (\ref{six})
can be straightforwardly done resulting in
\begin{equation}\label{seven}
\hspace*{-0.4cm}{\cal Z}(\lambda) = \int\limits_{-\infty}^{+\infty} \hspace*{-0.1cm} d \xi
\int\limits_{-\infty}^{+\infty} \hspace*{-0.1cm}
\frac{d \eta}{\sqrt{2 \pi}} ~
\frac{  \textstyle e^{ - i \eta \xi }  }{{\textstyle \lambda - i\eta ~-~{\cal F}(\xi,\lambda - i\eta)}}~,
\end{equation}

\vspace*{-0.3cm}

\noindent
where the function ${\cal F}(\xi,\tilde\lambda)$ is defined as follows
\begin{equation}\label{eight}
{\cal F}(\xi,\tilde\lambda) = \sum\limits_{k=1}^{ K_{max}(\xi) } 
\left[  \frac{ z_k^+}{s_1} + \frac{ z_k^-}{s_1} \right]
~e^{ - k\,s_1\tilde\lambda  }\,.
\end{equation}
As usual, in order to find the GCSP by  the inverse Laplace transformation,
it is necessary to study the structure of singularities of the  partition (\ref{seven}).
Since the HDM requires the fragment volume conservation, hereafter we will call
(\ref{seven}) as an isochoric partition or isochoric ensemble.

\vspace*{-0.1cm}


\section{Isochoric Ensemble Singularities}

\vspace*{-0.2cm}

For finite fragment surface 
the structure of  singularities of the isochoric partition (\ref{seven}) 
can be complicated. 
To see this let us first make the inverse Laplace transform:
\begin{eqnarray}\label{nine}
&&\hspace*{-0.6cm} Z(S_A)~ =
\int\limits_{\chi - i\infty}^{\chi + i\infty}
\frac{ d \lambda}{2 \pi i} ~  {\cal Z}(\lambda)~ e^{\textstyle   \lambda \, S_A } =
\nonumber \\
&&\hspace*{-0.6cm}\int\limits_{-\infty}^{+\infty} \hspace*{-0.1cm} d \xi
\int\limits_{-\infty}^{+\infty} \hspace*{-0.1cm}  \frac{d \eta}{\sqrt{2 \pi}}
\hspace*{-0.1cm} \int\limits_{\chi - i\infty}^{\chi + i\infty}
\hspace*{-0.1cm} \frac{ d \lambda}{2 \pi i}~
\frac{\textstyle e^{ \lambda \, S_A - i \eta \xi } }{{\textstyle \lambda - i\eta ~-~{\cal F}(\xi,\lambda - i\eta)}}~=
\nonumber \\
&&\hspace*{-0.6cm}\int\limits_{-\infty}^{+\infty} \hspace*{-0.1cm} d \xi
\int\limits_{-\infty}^{+\infty} \hspace*{-0.1cm}  \frac{d \eta}{\sqrt{2 \pi}}
\,{\textstyle e^{  i \eta (S_A - \xi)  } } \hspace*{-0.1cm} \sum_{\{\tilde\lambda_n\}}
e^{\textstyle  \tilde\lambda_n\, S_A }\hspace*{-0.0cm}
{\textstyle 
\left[1 - \frac{\partial {\cal F}(\xi,\tilde\lambda_n)}{\partial \tilde\lambda_n} \right]^{-1} \hspace*{-0.2cm},
}
\end{eqnarray}
where the contour  integral in $\lambda$ 
is reduced to the sum over the residues of all singular points
$ \lambda = \tilde\lambda_n + i \eta$ with $n = 0, 1, 2,..$, since this  contour 
in the complex $\lambda$-plane  obeys the
inequality $\chi > \max(Re \{  \tilde\lambda_n \})$.
Now all integrations in (\ref{nine}) can be done, and the GCSP acquires the form 
\begin{equation}\label{ten}
 Z(S_A)~ = \sum_{\{\tilde\lambda_n\}}
e^{\textstyle  \tilde\lambda_n\, S_A }
\left[1 - \frac{\partial {\cal F}(S_A,\tilde\lambda_n)}{\partial \tilde\lambda_n} \right]^{-1} \,,
\end{equation}
i.e. the double integral in (\ref{nine}) simply  reduces to the substitution   $\xi \rightarrow S_A$ in
the sum over singularities.
This remarkable answer is a partial example  of the general theorem on the Laplace-Fourier transformation
properties proved in \cite{Bugaev:04}. 

The simple poles in (\ref{nine}) are defined by the condition 
$\tilde\lambda_n~ = ~{\cal F}(S_A,\tilde\lambda_n)$ and the latter can be cast
as a system of two coupled transcendental equations
\begin{eqnarray}\label{eleven}
&&\hspace*{-0.2cm} R_n = ~  \sum\limits_{k=1}^{K_{max}(S_A) }  \left[ z_k^+ + z_k^- \right]
~{\textstyle e^{- k\,R_n} } \cos(I_n  k)\,,
\\
\label{twelve}
&&\hspace*{-0.2cm} I_n = -  \sum\limits_{k=1}^{K_{max}(S_A) } \left[ z_k^+ + z_k^- \right]
~{\textstyle e^{-k\,R_n} } \sin(I_n  k)\,, 
\end{eqnarray}
for dimensionless variables $R_n = s_1 Re(\tilde\lambda_n)$ and $I_n = s_1 Im(\tilde\lambda_n)$. 
So far Eqs. (\ref{eleven}) and (\ref{twelve}) are rather general and can be used for
particular models which specify the height of hills and depth of dales. 
But we can give an absolute supremum for the real root $(R_0; I_0 = 0)$ of these equations.  
For this purpose it is sufficient to  consider the  limit $K_{max}(S_A) \rightarrow \infty$,
because for $I_n = I_0 = 0$ the right hand side (r.h.s.) of (\ref{eleven}) is a monotonously
increasing function of $K_{max}(S_A)$.
Since $z_k^+ = z_k^- $ are the monotonously decreasing functions of $H_k$, the maximal value of 
the r.h.s. of (\ref{eleven})  corresponds to the limit of infinitesimally small amplitudes 
of deformations, $H_k \rightarrow 0$.
Under these conditions Eq. (\ref{twelve}) for $I_n = I_0 = 0$ becomes an identity and  
Eq. (\ref{eleven}) acquires the form 
\begin{equation}\label{thirteen}
R_0 \rightarrow ~  2 \sum\limits_{k=1}^{\infty } e^{- \frac{ \sigma_{\rm o} P_k H_k}{2 T s_1} }  
~{\textstyle e^{ - k\,R_0} }  = 2 \left[ e^{ R_0} - 1 \right]^{-1} \,,
\end{equation}
and we have $R_0 = s_1 \tilde\lambda_0 \approx 1.06009$. 
Since for $I_n \neq 0$ defined by (\ref{twelve}) the inequality $\cos(I_n k) \le 1$
cannot become the equality for all values of $k$ simultaneously, then 
it follows that the real root of (\ref{eleven}) obeys the inequality $R_0 > R_{n > 0}$.    
The last result means that in the limit of infinite fragment, $S_A \rightarrow \infty$, 
the GCSP is represented by the farthest right singularity among all simple poles $\{\tilde\lambda_n\}$ 
\begin{equation}\label{fourteen}
Z(S_A)\biggl|_{S_A \rightarrow \infty} ~ \approx
\frac{e^{\textstyle  \frac{R_0\, S_A}{s_1} } }{1 + \frac{R_0^2}{2} } 
\approx 0.6402~ e^{\textstyle  \frac{R_0\, S_A}{s_1} }
%
%
%
%
\end{equation}
There are two remarkable facts regarding (\ref{fourteen}): first, this result is model independent because
in the limit of vanishing amplitude of deformations all specific parameters of the model have dropped out; 
second, in evaluating (\ref{fourteen}) we did not specify the shape of the fragment under consideration,
but only implicitly required that the fragment surface together with deformations is a regular surface
without  self-intersections.  
Therefore, for vanishing amplitude of deformations the latter means that Eq. (\ref{fourteen})
should be valid for any self-non-intersecting surfaces.  

For  spherical fragments the r.h.s. of (\ref{fourteen}) becomes familiar, 
$ 0.6402~ e^{\textstyle  1.06009\, A^{2/3}  } $, which, combined with the Boltzmann factor of
the surface energy $e^{\textstyle - \sigma_{\rm o} A^{2/3}/T  } $, generates
the { following temperature dependent surface tension} of the large fragment 
\begin{equation}\label{fifteen}
\sigma (T) = \sigma_{\rm o} \left[ 1 - 1.06009 \frac{T}{\sigma_{\rm o} } \right] 
\end{equation}

%
%
\begin{figure}[ht]
\includegraphics[width=8.6cm,height=8.6cm]{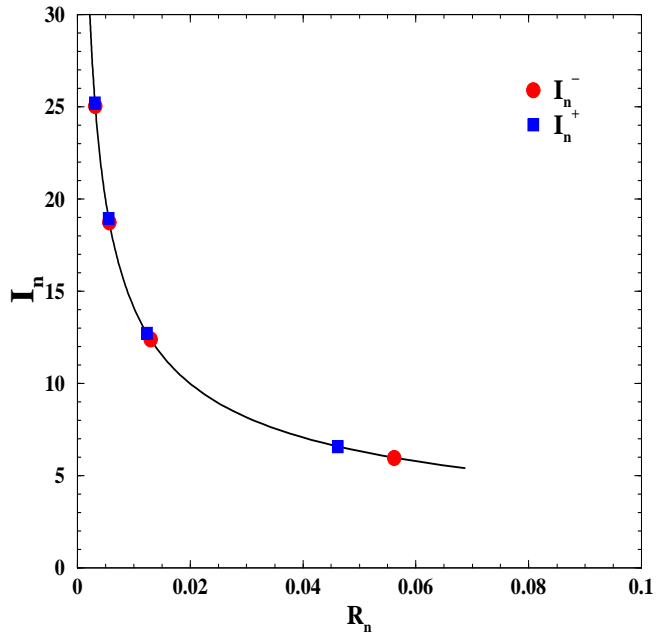}
\caption{
The first quadrant of the complex plane $s_1\tilde\lambda_n \equiv R_n + i I_n$  of the roots
of
the system of (\ref{sixteen}) and (\ref{seventeen}).
The circles and squares  represent the two branches $I_n^-$ and $I_n^+$ of the roots,
respectively. The curve is defined by the approximation given by
(\ref{nineteen}) and  (\ref{twenty}) (see text for more details).
}
  \label{fig1}
\end{figure}

\vspace*{-0.3cm}

\noindent
which means that the actual critical temperature of the three dimensional Fisher model should be
$T_c = \sigma_{\rm o}/ 1.06009$, i.e. 6 \% smaller than Fisher originally supposed.
{ Note please that this equation for the critical temperature 
 remains valid for the   temperature dependent $\sigma_{\rm o}$ as well.} 
It is also surprising that the degeneracy factor $0.6402$ in (\ref{fourteen})
is  only 12.5 \% larger than the corresponding factor of the self-avoiding polygons on
the two-dimensional square lattice \cite{SAP}.

For the large, but finite fragments it is necessary to take into account not only the farthest
right singularity $\tilde\lambda_0 = R_0/s_1$ in (\ref{ten}),  but all other roots with positive real part  
$R_{n>0} > 0$. In this case for each $R_{n>0}$ there are two roots $\pm I_n$ of (\ref{twelve}) because
the GCSP is real by definition.  
The roots of Eqs. (\ref{eleven}) and (\ref{twelve}) with largest real part are very insensitive
to the large values of $K_{max} (S_A)$, therefore, it is sufficiently good to  keep 
$K_{max} (S_A) \rightarrow \infty$.
Then for limit of vanishing amplitude of deformations Eqs. (\ref{eleven}) and (\ref{twelve}) can be,
respectively, rewritten as 
\begin{eqnarray}\label{sixteen}
&&\hspace*{-0.2cm}\frac{2 R_n}{R_n^2 + I_n^2} = 
~~{\textstyle e^{R_n} } \cos(I_n ) - 1\,,
\\
\label{seventeen}
&&\hspace*{-0.2cm}\frac{2 I_n}{R_n^2 + I_n^2} = -
~{\textstyle e^{R_n} } \sin(I_n )\,.
\end{eqnarray}
After some algebra the system of (\ref{sixteen}) and (\ref{seventeen})
can be identically reduced to a single equation for $R_n$
\begin{equation}\label{eighteen}
\cos\left( {\textstyle \left[ \frac{ 4 (1 + R_n)}{e^{2 R_n} - 1} - R_n^2 \right]^{1/2} } \right) = \cosh R_n - 
\frac{\sinh R_n}{1 + R_n}\,, 
\end{equation}
and  the  quadrature 
$I_n^2 = \sqrt{\frac{ 4 (1 + R_n)}{e^{2 R_n} - 1} - R_n^2}$.
The analysis shows that besides the 
opposite signs there are two branches of solutions, $I^+_{n}$ and $I^-_{n}$, 
for the same $n  \ge 1$ value: 
\begin{eqnarray}\label{nineteen}
|I^\pm_{n}| & \approx & 2 \pi n  \pm \frac{1}{\pi n}\,, 
\\
\label{twenty}
R_{n} & \approx & \pi^2 n^2 + 1 - \pi n \sqrt{ \pi^2 n^2 + 2}\,.
\end{eqnarray}
The exact solutions $(R_n ; I^\pm_n)$ for $ n > 1$ which have the largest real part  are shown in Fig. 1.
together with the curve  parametrized by functions $I_x^+$ and $R_x$
taken  from Eqs. (\ref{nineteen}) and (\ref{twenty}),
respectively.
From Eq. (\ref{twenty}) and Fig. 1. it is clearly seen that the  largest real part $R_1 \approx 0.0582$ is 
about 18 times smaller than $R_0$, and, therefore, already for the fragment of a few nucleons   the correction 
to the leading term (\ref{fourteen}) is exponentially small.
Using the approximations (\ref{nineteen}) and (\ref{twenty}),
for $n > 2$ one can  estimate  
\begin{equation}\label{twone}
 \biggl| {\textstyle  e^{\tilde\lambda_n\, S_A} 
\left[1 - \frac{\partial {\cal F}(S_A,\tilde\lambda_n)}{\partial \tilde\lambda_n} \right]^{-1} }\biggr|  \le
e^{\textstyle  \frac{S_A}{ 2 \pi^2 n^2 s_1} }~ /~ (2 \pi^2 n^2)  
\end{equation}
the upper limit of
the $(R_n ; I^\pm_n)$ root contribution into the GCSP (\ref{ten}).
This result shows that the total contribution of all complex poles in (\ref{ten})
is negligibly small compared to the leading term (\ref{fourteen}) for a fragment of a few nucleons already.
The latter, however, requires a more careful accounting for the volume conservation of a fragment.



\section{Concluding Remarks}

The developed model  allows us to give the upper limit for the surface entropy because
it corresponds to the vanishing amplitude of deformations and, therefore,
the specific features of the model were irrelevant for our analysis. 
To find the next order correction  to  the surface entropy one has already to consider
the underlying   model for deformations.  
We, however, would like to show how the power law may arise within the HDM. 
For this purpose let us consider the left equality in (\ref{thirteen}) which is valid 
for small heights of deformations. It can be shown that the following ansatz $(S_A >> s_1)$
for the deformation energy
\begin{equation}\label{twtwo}
\frac{ \sigma_{\rm o} P_k H_k}{2 T s_1}  \rightarrow - \frac{3}{2}~ k~ \tau 
{\textstyle \left[1 + \frac{2}{R_0 (2 + R_0)} \right] }~ 
 \frac{s_1}{S_A} \ln \left( \frac{s_1}{S_A} \right)
\end{equation}
of $k$ nucleon base, 
indeed, generates the Fisher power law $A^{-\tau}$ for the GCSP (\ref{fourteen}) of { an  $A$-nucleon} fragment. 
From (\ref{twtwo}) it is clearly seen that the term $- k \frac{s_1}{S_A} \ln \left( \frac{s_1}{S_A} \right) $
is the entropy  which gives an {\it a priori} uncertainty to measure the position of $k$ nucleons of area $s_1$ 
on the surface of the fragment. 
A comparison of (\ref{twtwo}) with any $k R_n  > 0$ in the left equality 
(\ref{thirteen})
 shows that in the limit $S_A >> s_1$ the ansatz
(\ref{twtwo}) corresponds to a negligible correction compared to the exponentials $e^{\frac{R_n S_A}{s_1} }$.
Therefore, the Fisher power law is  { too  delicate} 
for the model  developed to study 
the surface partition of large fragments. 

In conclusion, we developed a simple statistical model which allowed us to derive analytically
the general expression (\ref{ten}) for the  GCSP. 
This result is achieved by the Laplace-Fourier transformation technique to the isochoric ensemble. 
We named this ensemble because the HDM obeys the
volume conservation of a fragment under consideration.  The volume conservation
is accounted for  
by the equal statistical probabilities for the hills and dales of the same base.
The limit of vanishing deformations allowed us to find a  supremum for the surface entropy of
the large fragments which, remarkably, exceeds the Fisher original assumption by only 6 percent.   
The analytical analysis of the corrections to the GCSP (\ref{fourteen})
originating from the complex roots of Eqs. 
(\ref{eleven}) and (\ref{twelve}) showed that these corrections are negligible already for
the fragment of a few nucleons.  
The HDM  allows one to study the statistical mechanics of 
volume deformations of finite or even small fragments,
but this task  requires further refinements of the model.

{\bf  Acknowledgments.}
This work was supported by the US Department of Energy.




\begin{thebibliography}{99}


\bibitem{Fisher:64}
M.E. Fisher, Physics {\bf 3} (1967) 255.


\bibitem{Moretto}
%
L. G. Moretto {\it et. al.}, Phys. Rep. {\bf 287} (1997) 249.

\bibitem{Ising:clust}
%
C.~M.~Mader {\it et al.},
Phys. Rev.  {\bf C 68} (2003) 064601.




\bibitem{Bondorf:95}
J. P. Bondorf {\it et al.},
Phys. Rep. {\bf 257} (1995) 131.


\bibitem{Bugaev:00}
K. A. Bugaev {\it et. al.},
Phys. Rev. {\bf C62} (2000) 044320;
arXiv:nucl-th/0007062 (2000).

\bibitem{Bugaev:01}
K. A. Bugaev {\it et. al.}, 
Phys. Lett. {\bf B 498} (2001) 144;
arXiv:nucl-th/0103075 (2001).

\bibitem{Reuter:01}
P.~T.~Reuter and K.~A.~Bugaev,
Phys.\ Lett.\ B {\bf 517} (2001) 233.



\bibitem{Goren:81}
M. I. Gorenstein, V. K. Petrov and G. M. Zinovjev, Phys. Lett.
{\bf B 106} (1981) 327.

\bibitem{Zeeb:02}
G.~Zeeb, K.~A.~Bugaev, P.~T.~Reuter and H.~Stocker,
arXiv:nucl-th/0209011.

\bibitem{Bugaev:04}
%
K.~A.~Bugaev, arXiv:nucl-th/0406033.

\bibitem{SAP}
%
I. Jensen and A. J. Guttmann, J. Phys. {\bf A 32} (1999) 4867.


\end{thebibliography}
\end{document}